\documentclass[sigconf,authorversion,nonacm,10pt]{acmart}
\usepackage[utf8x]{inputenc}
\usepackage[T1]{fontenc}             
\usepackage{lmodern}
\usepackage[english]{babel}
\usepackage{amsmath}
\usepackage[procnames]{listings}
\usepackage{color}
\usepackage{caption}
\usepackage{subcaption}
\usepackage{array,graphicx}
\usepackage[squaren]{SIunits}
\usepackage[abbreviations]{foreign}
\usepackage{algorithm}
\usepackage[noend]{algpseudocode}
\usepackage{setspace}
\usepackage{pifont}
\usepackage{listings}
\usepackage{multirow}
\usepackage{stmaryrd}      
\setcopyright{acmcopyright}
\newcommand{\sys}{{rgpdOS}\xspace}

\begin{document}

\title{rgpdOS: GDPR Enforcement By The Operating System}

\author{Alain Tchana}
\affiliation{%
  \institution{ENS Lyon}
  \city{Lyon}
  \state{France}
}
\email{alain.tchana@ens-lyon.fr}

\author{Raphael Colin}
\affiliation{%
  \institution{ENS Lyon}
  \city{Lyon}
  \state{France}
}
\email{raphael.colin@ens-lyon.fr}

\author{Adrien Le Berre}
\affiliation{%
  \institution{ENS Lyon}
  \city{Lyon}
  \state{France}
}
\email{adrien.leberre@ens-lyon.fr}

\author{Vincent Berger}
\affiliation{%
  \institution{University of Lyon 3}
  \city{Lyon}
  \state{France}
}
\email{vincent.berger@univ-lyon3.fr}

\author{Benoit Combemale}
\affiliation{%
  \institution{University of Rennes}
  \city{Rennes}
  \state{France}
}
\email{benoit.combemale@irisa.fr}

\author{Natacha Crooks}
\affiliation{%
  \institution{Berkeley}
  \city{California}
  \state{USA}
}
\email{ncrooks@berkeley.edu}

\author{Ludovic Pailler}
\affiliation{%
  \institution{University of Lyon 3}
  \city{Lyon}
  \state{France}
}
\email{ludovic.pailler@univ-lyon3.fr}

\renewcommand{\shortauthors}{Alain Tchana et al.}

\begin{abstract}
The General Data Protection Regulation (GDPR) forces IT companies to comply with a number of principles when dealing with European citizens' personal data.
Non-compliant companies are exposed to penalties which may represent up to 4\% of their turnover.
Currently, it is very hard for companies driven by personal data to make their applications GDPR-compliant, especially if those applications were developed before the GDPR was established.
We present \sys, a GDPR-aware operating system that aims to bring GDPR-compliance to every application, while requiring minimal changes to application code.
\end{abstract}



\maketitle

\sloppy

\section{Introduction}
\label{introduction}
\begin{figure}
     \centering
     \begin{subfigure}[b]{0.22\textwidth}
         \centering
         \includegraphics[width=\textwidth]{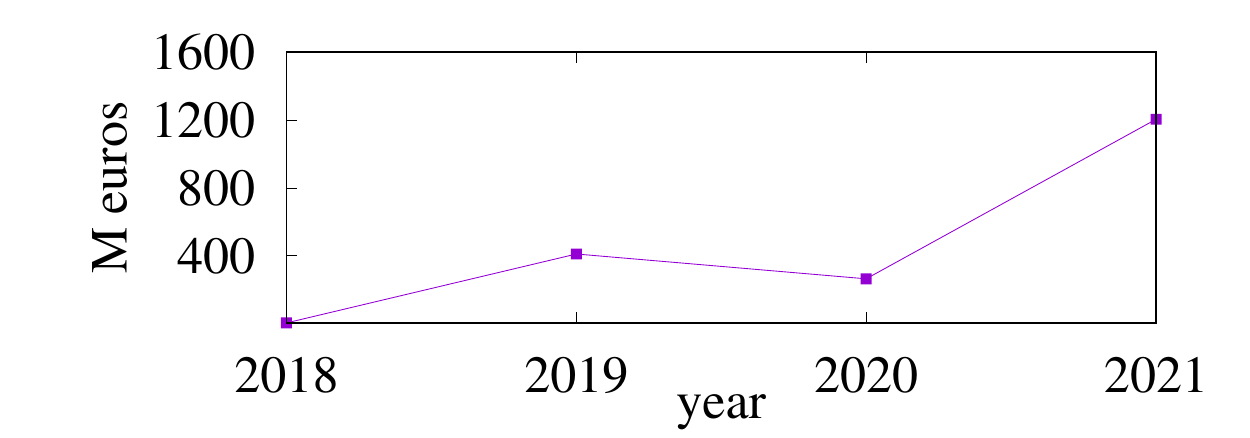}
     \end{subfigure}
     \begin{subfigure}[b]{0.22\textwidth}
         \centering
         \includegraphics[width=\textwidth]{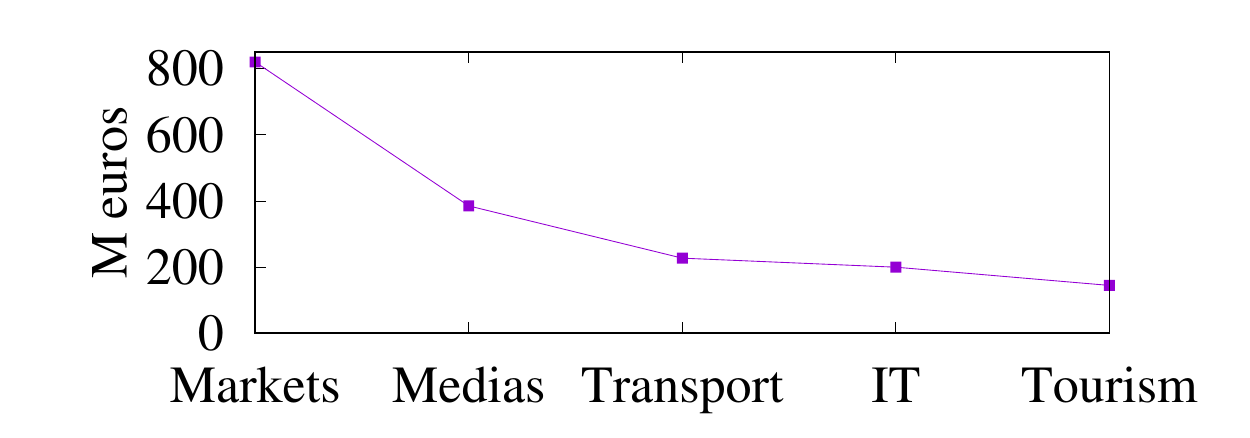}
     \end{subfigure}
        \caption{Relying on~\cite{stats-rgpd}: (left) total amount of penalties; (right) top 5 most sanctioned business sectors.}
        \label{fig:gdpr-sanctions}
\end{figure}

The General Data Protection Regulation (GDPR), introduced in May 2018, forces data operators to comply with a number of principles made to enforce personal data (PD) protection for European citizens (hereafter the subjects).
Data operators failing to comply with the regulation can be fined up to 4\% of their global turnover.
For example, in March of 2022, the Irish Data Protection Commission sentenced Meta to a fine of €17M~\cite{dpc}.
As summarized in Fig.~\ref{fig:gdpr-sanctions} left, the amount of penalties imposed to companies increases every year, topping 1.2 billion euros in 2021.
Also, as shown in Fig.~\ref{fig:gdpr-sanctions} right, companies of all types are impacted.
For instance, in 2020 the CNIL in France penalized two doctors (€9K) for hosting medical images on a server which was freely accessible on the Internet~\cite{cnil}.

\begin{figure}
     \centering
         \includegraphics[width=.50\columnwidth]{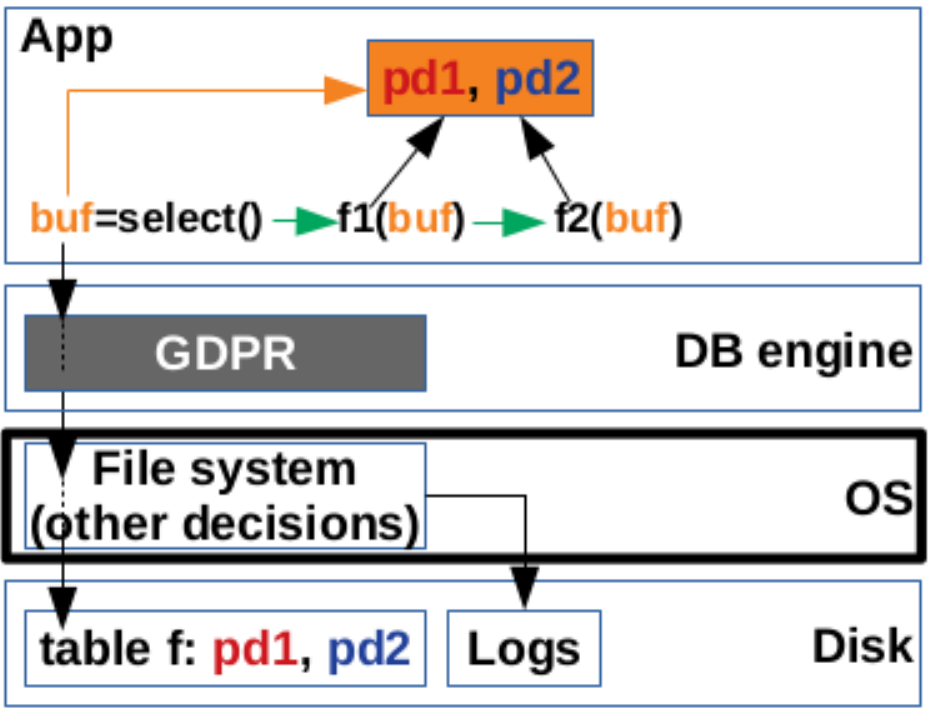}
         \caption{GDPR at the DB engine level in userspace.}
         \label{fig:db_level}
\end{figure}
However, bringing computer science and law together is not an easy task~\cite{catala1, catala2}.
The law mainly targets the new opportunities and dangers that information technologies create.
Therefore, it is necessary to rethink applications for GDPR compliance, while maintaining usability, performance and velocity.
Very few studies investigated GDPR compliance at application design time.
Among existing works, some~\cite{pbd1,pbd2} suggested guidelines, mainly in terms of documents which provide best practices.
However, these efforts are lacking a true operational form.
To our knowledge, Shastri et al.~\cite{Shastri_2020}, Schwarzkopf et al.~\cite{schwarzkopf.lecture.2019} and Istvan et al.~\cite{Zsolt} are the only operational works.
The two former take into account the GDPR within the application's data base (DB) engine in userspace, thus relying on a general purpose OS, as shown in Fig.~\ref{fig:db_level}.
This approach has two main limitations:
first, it is application-specific;
second, the OS's filesystem upon which the DB engine is running can take actions that contradict the latter, thus potentially violating GDPR compliance.
For illustration, the filesystem's logging mechanism can compromise the GDPR's \textit{right to be forgotten} as data deleted by the DB engine can still be present in the filesystem's logs.
Istvan et al.~\cite{Zsolt} studied how existing storage hardware technologies can be leveraged by applications to ensure the RGPD.
Here, the developer deals with the RGPD, which is far from her expertises.

\begin{figure}
     \centering
         \includegraphics[width=.8\columnwidth]{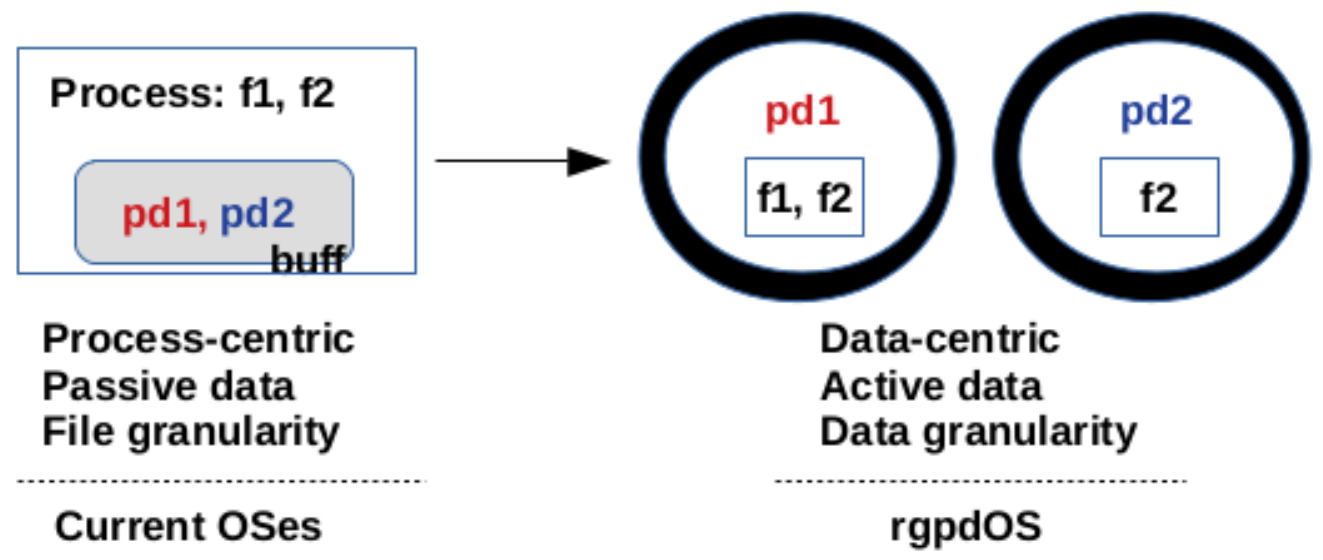}
         \caption{Our vision}
         \label{fig:rgpdos_level}
\end{figure}

In this paper introduces \sys, a fundamentally new approach to make applications GDPR compliant.
\sys mainly proposes a \emph{GDPR-aware operating system (OS)} for two main reasons.
First, by working at the OS level, any application can benefit from GDPR compliance ensured by \sys.
Second, we want to build an end-to-end solution that enforces GDPR at any software level.
This is contrary to existing solutions which need to trust the huge code base of existing OSes (e.g., Linux has about 26M LOC).
As a consequence, one of our main goal is to keep \sys as \textit{lightweight} as possible.

\sys acts as a framework which forces the data operator to respect a number of \emph{technical} rules, which in turn allows the OS to ensure GDPR compliance.
This way, the data operator does not need to deal with the articles of the Law, but only technical rules, which are easier to comply with according to their background and more straightforward.
According to the European Union Agency for Network and Information Security (ENISA)'s definition, \sys can be classified as a Privacy Enhancing Technology (PET)~\cite{pet}.
Using \sys a data operator is demonstrating a conscious effort towards GDPR compliance like imposed by its 25th article.
In addition, this certainly helps to reassure subjects about the respect of their rights.

We build \sys following three new principles which go against current OS architecture, summarized in Figure~\ref{fig:rgpdos_level}.
(Idea~1) In \sys, we introduce the abstraction of \textit{active data}, unlike current OSes in which data is \textit{passive}.
Active data is able to enforce the respect of consents defined by its subject.
This is the role of the black membrane depicted in Fig.~\ref{fig:rgpdos_level} right.
(Idea~2) \sys is \textit{data-centric}, unlike current OSes which are \textit{process-centric}.
In the latter, the process brings data to its domain (virtual address space), as illustrated in Fig.~\ref{fig:db_level} (see $pd1$ and $pd2$).
A function which should not access some PD (\textit{e.g.}, because the PD's subject did not approve this usage) could still gain access to them (e.g., accidentally due to a use-after-free~\cite{uaf} vulnerability).
Fig.~\ref{fig:db_level} illustrates such a situation where function $f2$ accidentally accesses $pd2$.
\sys reverses this power balance and runs the function in the \emph{PD's domain} as shown in Fig.~\ref{fig:rgpdos_level} right.
(Idea~3) In \sys, we advocate for a structured database-oriented filesystem, unlike current OSes such as Linux which are file-based.
The latter is coarse grained and not aware of the notion of \emph{data} which can have different types.
The GDPR requires systems to work at the granularity of individual PD pieces while traditional file systems see files, which are just bytes without particular meaning.

In our vision, the same server should still be able to process PD and NPD sequentially or at the same time (e.g., a single application may need to combine these two data types).
Each data type should be processed using a distinct OS as we want to keep \sys as lightweight and simple as possible, dedicated to PD processing.
To achive this goal, we introduce \emph{purpose kernel}, a new kernel model that we follow to design the entire machine kernel.

To summarize, from the Systems research perspective, this project will investigate three novel research concepts: data centric, active data, database-oriented file system, and purpose kernel model.
In this paper, we are laying the groundwork around these concepts, which raise several challenges.
We do not list all of the latter in this paper as we are still discovering them.

The rest of this paper is organized as follows: \S~\ref{design} presents the design of \sys and the main implementation choices are presented in \S~\ref{implementation}, \S~\ref{illustration} shows how \sys can effectively enforce some example GDPR rules, finally \S~\ref{conclusion} presents the conclusion.
\section{Design}
\label{design}
\begin{figure}[!h]
    \centering
        \includegraphics[width=.6\columnwidth]{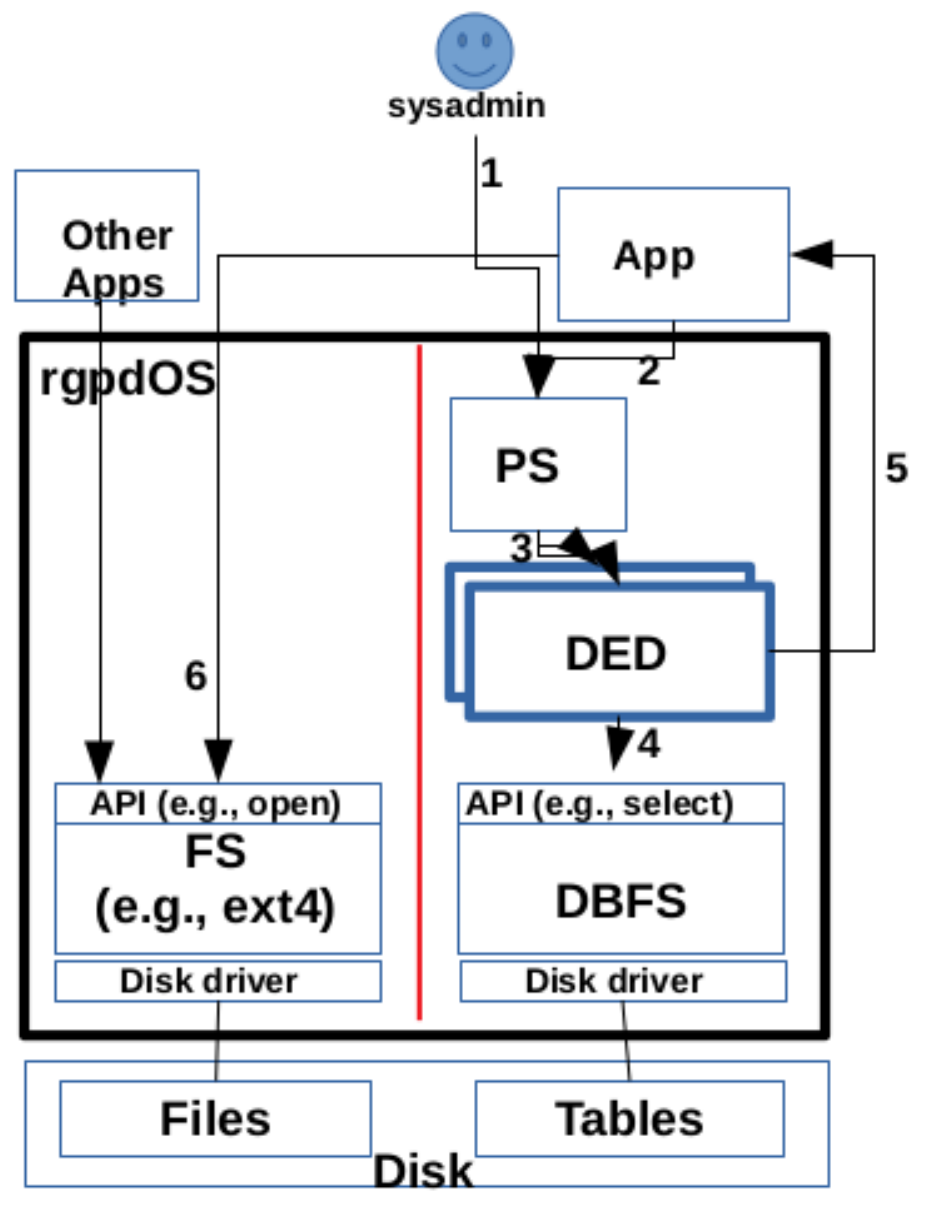}%
        \includegraphics[width=.4\columnwidth]{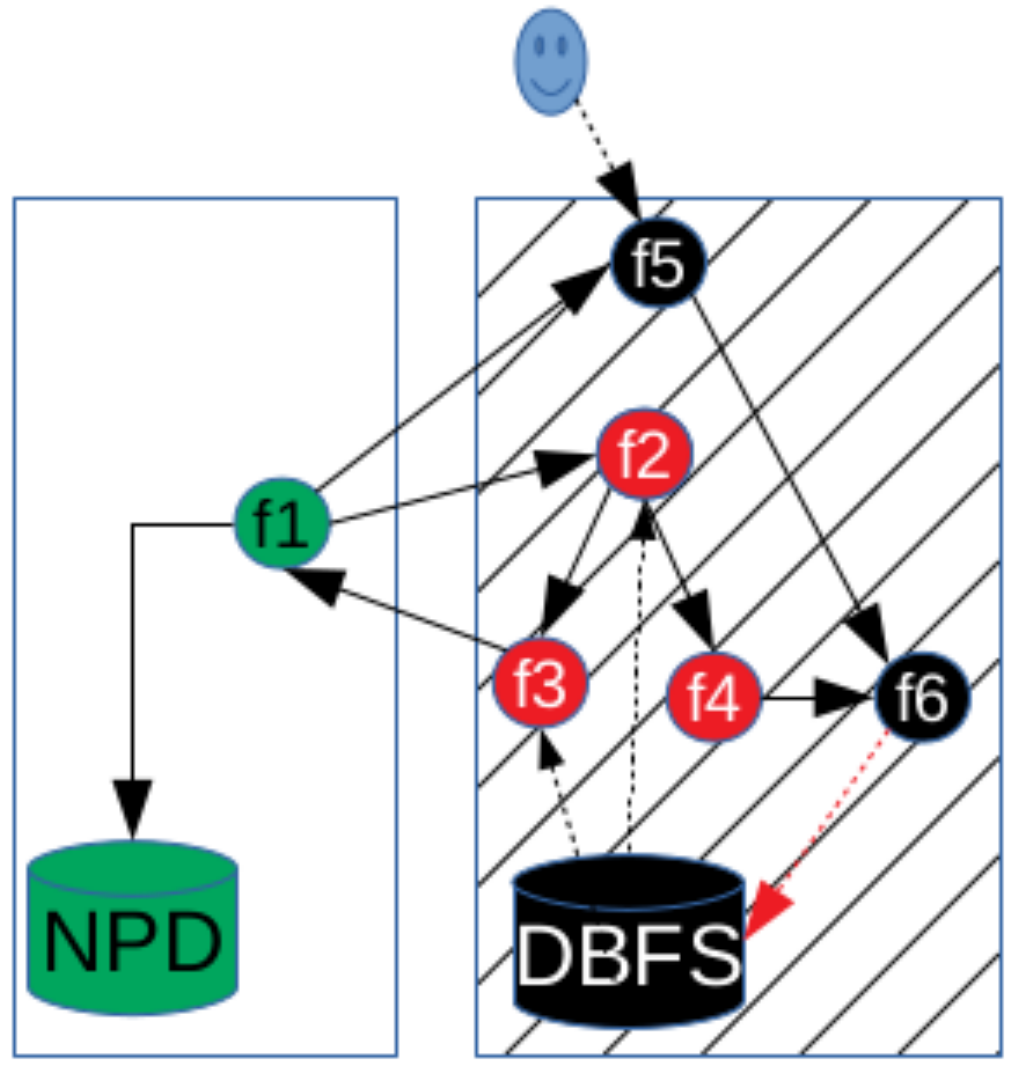}%
        
    \caption{(left) Architecture of \sys, followed by (right) its programming model and execution.}
    \label{fig:architecture}
\end{figure}

The main guiding principle of \sys is as follows: \emph{every access to PD must be controlled by \sys}.
\sys therefore intervenes during the entire life cycle of each piece of PD:
from collection to processing and destruction.
Fig.~\ref{fig:architecture} left shows the architecture of \sys, described below.

\paragraph{\textbf{Purpose kernel model}}
In this model, the kernel is the agregation of several sub-kernels where each sub-kernel achieves a specific purpose.
The sub-kernels can be organized into three categories:
\begin{itemize}
	\item IO driver kernels: every IO device is managed by a dedicated kernel which is mainly composed of the device driver. They are lightweight kernels.
	\item a general purpose kernel: it hosts and processes NPD. In addition, it does not include IO drivers.
	\item \sys, a PD GDPR-aware kernel: it hosts and processes PD.
\end{itemize}
The different kernels cooperate to (dynamically) partition CPU and memory resources.
We remove the IO devices under the management of the general purpose kernel because they are traversed by PD.
Therefore, \sys and IO driver kernels are the systems that we plan to prove in order to guarantee end-to-end GDPR compliance.

\paragraph{\textbf{File System}}
\sys contains at least two filesystems:
the first one named DBFS is based on a structured DB and is used to store PD.
Every PD has a precise \emph{type} which corresponds to a table in the kernel's DB.
In \sys, data types must be created in DBFS prior to use by a data processing application.
PD structuration (typing) is important for several reasons.
First, it allows to apply the GDPR on different \emph{views} of the same PD (see below).
Second, it enables subjects to obtain their PD from the data operator in a structured and machine-readable format like the GDPR prescribes (see \S~\ref{illustration}).
Using a DB as a filesystem is appropriate in this case, especially since DB engines have seen significant improvement over the last years\cite{dbos}.

The second filesystem in \sys is used to store non-personal data.
It can be implemented with a traditional filesystem (e.g., \emph{ext4}) which works at the file granularity.

DBFS can only be accessed through the components of \sys.
In other words, DBFS is not visible from the outside and every direct access attempt from the outside is blocked by using a security mechanism (e.g., Linux Security Module).
As for the second filesystem, it is accessible by any process running on the machine.

\paragraph{\textbf{PD membrane}}
This is the first demonstration of the notion of \emph{active data}.
Each PD stored in DBFS includes a \emph{membrane}.
The configuration of the membrane and PD types is handled by the data operator (hereafter sysadmin).
The membrane features different categories of metadata, among the most important ones are:
the origin of the PD (it could be the subject, the sysadmin, or another data operator);
consents relative to each data processing operation;
time to live;
level of sensibility;
the interface to use for data collection (e.g., a web page to gather data directly from the subject) when the data is not present in DBFS; 
etc.
Consent indicates the permissions of each operation on PD.
A data type \emph{view} corresponds to a specific representation or fragment of the data type.
This notion enables our system to implement the GDPR's \emph{data minimization} principle.
About the sensibility level, it is useful since the GDPR prescribes that sensitive data (e.g., a social security number) be stored separately from less sensitive data (\textit{e.g.} a name).
The time to live is directly requested by the GDPR and can be used to implement the \textit{right to be forgotten}.
Finally, the data collection interface will be used by \sys to initialize DBFS with data collected from the appropriate data sources.

In order to illustrate how \sys works, listing~\ref{lst:exampletype} shows a simplified example of a data type declaration (\emph{user}), as well as its default membrane.
This example also defines two views \textit{v\_name} and \textit{v\_anon}.
The former makes only the name of a user visible while the latter view only allows access to the user's age.
The \emph{consent} keyword indicates the default consent to apply when data of this type is created (collected).
The GDPR allows PD processing when the latter relies on a legitimate basis, for example, the user's consent.

The default consent allows the data operator to express operations that are backed by a legitimate basis, and thus do not need the specific subject's consent.
In our example, operation \emph{purpose1} can access the entirety of the \emph{user} data type,
\emph{purpose3} can only manipulate the age,
and \emph{purpose2} is not allowed to process this data type at all.

\definecolor{keywords}{RGB}{255,0,90}
\definecolor{comments}{RGB}{0,0,113}
\definecolor{red}{RGB}{160,0,0}
\definecolor{green}{RGB}{0,150,0}
\lstset{language=C, 
        basicstyle=\footnotesize,
	    numbers=left, 
        keywordstyle=\color{keywords},
        commentstyle=\color{comments},
        stringstyle=\color{red},
        showstringspaces=false,
        identifierstyle=\color{green}}    

\begin{lstlisting}[caption=Example of a personal data definition type,label={lst:exampletype},frame=tlrb]{Name}
type user {
  fields {
    name: string,
    pwd: string,
    year_of_birthdate: int
  };  
  view v_name {
    name
  };
  view v_ano {
    age
  };   
  consent {
    purpose1: all,
    purpose2: none,
    purpose3: ano
  };
  collection {
    web_form: user_form.html,
    third_party: fetch_data.py
  };
  origin: subject;
  age: 1Y;
  sensitivity: hight;
}
\end{lstlisting}

\paragraph{\textbf{Programming model}}
\sys dictates that applications willing to process PD be programmed as a workflow of functions, as shown in Fig.~\ref{fig:architecture} right.
Functions can be organized into two categories:
($F_{npd}$) functions that do not access PD (f1 in Fig.~\ref{fig:architecture} right)
and ($F_{pd}$) functions that need to access PD (f2-f6 in Fig.~\ref{fig:architecture} right).
$F_{npd}$ can be implemented with a single function, as shown in Fig.~\ref{fig:architecture} right.
$F_{npd}$ is split into two subcategories:
($F^w_{pd}$) functions that modify the state of DBFS (f5 and f6 in Fig.~\ref{fig:architecture} right)
and ($F^r_{pd}$) functions that do not (f2-f4 in Fig.~\ref{fig:architecture} right).
$F^w_{pd}$ functions are natively provided by \sys (they are \emph{built-in}) whereas $F^r_{pd}$ functions are developed by the data operator.
Built-in functions (described below) ensure that every PD is correctly wrapped, that is it always includes a membrane.
Among built-in functions, we can list \texttt{update}, \texttt{delete}, \texttt{copy} and \texttt{acquisition} (for data collection).
$F^r_{pd}$ functions are forbidden to make syscalls that could leak PD (e.g., \texttt{write}).

Every $F_{pd}$ function is the implementation of a unique data processing \emph{purpose}.
Functions can be implemented in any programming language (e.g., C).
As for the purpose declaration, we introduce a new very high level language as purposes should probably be written by the project manager while the actual implementation is written by developers.
In \sys's jargon, we call \emph{data processing} a pair composed of one \textit{purpose} and its \textit{implementation}.
Listing~\ref{lst:finalite} shows an example where the implementation is done in C and the purpose is denoted by a C comment.
In this example, this processing's goal is to compute the age of the input user.
Since subjects can restrict each purpose to a limited set of PD's views, the implementation must be aware that some fields of the data type might not be
available, as illustrated in line 4 of listing~\ref{lst:finalite}.
In the latter, the computation is only possible for users for which the subject authorized access to the \textit{age} field for \textit{purpose3}.

\definecolor{keywords}{RGB}{255,0,90}
\definecolor{comments}{RGB}{0,0,113}
\definecolor{red}{RGB}{160,0,0}
\definecolor{green}{RGB}{0,150,0}
\lstset{language=C, 
        basicstyle=\footnotesize, 
        keywordstyle=\color{keywords},
        commentstyle=\color{comments},
        stringstyle=\color{red},
        numbers=left,
        firstnumber=1,
        showstringspaces=false,
        identifierstyle=\color{green}}
\noindent\begin{minipage}{.5\textwidth}
\begin{lstlisting}[caption=Example of data processing code,label={lst:finalite},frame=tlrb]{Name}
#include "/etc/rgpdos/ps/types.h"
/* purpose3 */
struct age_pd compute_age(struct user_pd user) {
  if (user.age) {//is age allowed to be seen?
    return current_year()-user.year_of_birthdate;
  }
  else {
    // error
  }
}
\end{lstlisting}
\end{minipage}\hfill
\begin{minipage}{.5\textwidth}
\begin{lstlisting}[caption=Example of main application,label={lst:main},breaklines=true,frame=tlrb]{Name}
#include "/etc/rgpdos/ps/ps.h"
int main() {
  int age = ps_invoke(modpol,ref,"compute_age", web_form, 0);
}
\end{lstlisting}
\end{minipage}

Any $F_{pd}$ function takes as input the identifier of a PD or a PD type, which will be used latter by \sys to automatically generate a request to extract data from DBFS.
Finally, when a $F_{pd}$ function wants to return some PD to the calling application, \sys instead returns a reference or ID.
Subsequently, the main application (app in Fig.~\ref{fig:architecture} left, f1 in Fig.~\ref{fig:architecture} right, or \texttt{main} in listing~\ref{lst:main}) never manipulates real PD within its address space.

\paragraph{\textbf{\sys built-in functions}}
In order to motivate \sys's built-in functions, let us consider three examples:
\begin{itemize}
	\item \texttt{copy}: \sys must ensure membrane consistency across all copies of the same PD.
	\item \texttt{collection}: \sys must ensure privacy and traceability (by keeping track of each PD's origin) from the moment PD enters the system.
To this end, \sys requests the needed metadata to fill the membrane with at data collection time.
Thanks to this requirement, each entry in DBFS is always correctly wrapped with its membrane.
However, \sys leaves the configuration of the collection interface (e.g., web form) to the data operator.
	\item \textit{delete}: \sys must ensure the GDPR's \textit{right to be forgotten}.
\end{itemize}

\paragraph{\textbf{Processing Store(PS)}}
This is the second component of \sys.
It is the only \sys entry point.
Its public interface consists of two functions:
\texttt{ps\_register} and \texttt{ps\_invoke}.
Every $F_{pd}$ function must be registered (\texttt{ps\_register}) first in PS before they can be invoked.
On call to \texttt{ps\_register}, PS makes the following checks:
if the function has no specified purpose, it is rejected;
if the specified purpose does not ``match'' with the corresponding implementation, PS raises an alert that requires an explicit sysadmin approval\footnote{Let's recall that the goal of \sys is to guide the data operator in the path of GDPR compliance.}.
PS is also the entry point for the invocation of $F_{pd}$ functions, through the \texttt{ps\_invoke} call, as shown in Fig.~\ref{fig:architecture} left and illustrated in listing~\ref{lst:main}.
\texttt{ps\_invoke} takes as input the reference of a data processing operation, optionally a reference to PD, a data collection method and a boolean indicating whether or not the data collection function is to be called to initialize DBFS.

\paragraph{\textbf{Data Execution Domain (DED)}}
This is the third and final component of \sys.
Any $F_{pd}$ function is always executed as an instance of the DED, an environment that ensures GDPR compliance on manipulated PD.
When PS receives a \texttt{ps\_invoke} call, it instantiates a DED, which in turn follows a number of steps.
(\texttt{ded\_type2req}) the DED translates the processing's input parameter type to requests at the destination of DBFS;
(\texttt{ded\_load\_membrane}) the first request actually submitted to DBFS is fetching the membrane of the involved PD;
(\texttt{ded\_filter}) the DED filters the queried membranes in order to only keep references to PD which approve the current processing;
(\texttt{ded\_load\_data}) the second request issued to DBFS fetches the actual data for PD which passed the previous filter;
(\texttt{ded\_execute}) the processing is executed on the fetched data;
(\texttt{ded\_build\_membrane}) if any PD is generated by the processing, it is assigned a membrane;
(\texttt{ded\_store}) if any PD is generated by the processing, it is stored in DBFS with its newly computed membrane;
(\texttt{ded\_return}) the DED returns to the caller non PD as well as references to PD.

\paragraph{\textbf{Enforcement}}
In order to guarantee any PD is only accessed under the supervision of \sys, the latter enforces a number of restrictions that we summarize here:
(1) PS is the only component that is able to access stored processings;
(2) PS is the only entry point to invoke a processing;
(3) every PD stored in DBFS must have a membrane attached to it;
(4) DED is the only component that is able to access DBFS directly.
\section{Implementation (in progress)}
\label{implementation}
This section presents the set of design choices that we make to implement \sys.

(1) Besides the intrinsic challenges raised by the new paradigms that \sys introduces, we impose on ourselves another challenge which is the implementation of \sys by revisiting Linux.
We do this to give \sys a chance to be easily adopted.
We are currently implementing \sys following the semi-microkernel model~\cite{10.1145/3477132.3483581,ghost}.
The latter runs some Linux kernel services as userspace processes for several reasons such as quick customization.
In \sys, DBFS, PS and DED run outside the Linux kernel while the second filesystem could be Linux's native filesystem.
To provide DBFS, we are rearchitecting uFS~\cite{10.1145/3477132.3483581} in order to implement a database oriented filesystem.
Notice that uFS is file based.
The only part of uFS that we keep is the implementation of the inode concept.
Starting from this point, PD in DBFS is represented by two major inode trees. 
The first tree gathers every PD from all subjects, with a separate set of inodes for each of them, grouping not only their personal data but also the membrane.
Moreover, the second major tree provides the database structure, with a core inode or subset of inodes for each table describing the structure of the contained data, the different fields of the table, and a list of subject's inodes, providing an easy link to quickly fetch the corresponding pieces of information.
Finally, a dedicated set of inodes describes the general structure of the data encoded in the inode subtree of each subject: meant to be accessed only once by the filesystem during a given live session.
This allows the system to properly format the pieces of data when returning them to DED.

(2) Concerning security enforcement, we rely on the Linux Security Module (LSM) framework.
After the analysis of the major LSM implementations, we observed that SELinux and Smack can do the job.
We leverage Linux Seccomp BPF to avoid functions which operate on PD to perform syscalls that can leak data.

(3) Different techniques can be used to ensure DED protection including TEEs like Intel SGX~\cite{sgx}.
Furthermore, DED could be executed in multiple locations with the help of \textit{Processing in Memory}~\cite{pim} (\textit{e.g.} UPMEM)
and \textit{Processing in Storage}.

(4) Finally, checking if a processing's implementation matches its purpose is a challenging problem which is not yet addressed in \sys.
We plan to investigate approaches borrowed from several research domains such as Semantic and AI.
\section{Illustration}
\label{illustration}
This section shows how \sys can ensure compliance with two example GDPR rules:
the \textit{right of access} and the \textit{right to be forgotten}.

\paragraph{\textbf{Right of access}}
The way that data operators are asked by the GDPR to send PD and processings to the subject is unclear.
As a matter of fact, the GDPR specifies that PD must be ``structured and machine-readable''.
Let's say for example that a company stores the first and last names of data subjects with key value pairs (\texttt{first\_name: "Chiraz", last\_name: "Benamor"}).
This pairing is structured, and is therefore machine readable, but what makes it exploitable is the fact that the keys make sense.
According to the GDPR, nothing is preventing this company from sending this same data in a structured, but inadequate way (\texttt{Chiraz: "Benamor"}).
Leveraging \sys, which enforces PD structuration, an official authority could simply ask that data be sent to subjects as it is stored in DBFS.
As for informing subjects about processings executed on their PD, this is easily obtained thanks to the DED, which logs every executed processing.
This log is organized so that it can give information about executed processings for each piece of PD.

\paragraph{\textbf{Right to be forgotten}}
This right is implemented as a built-in function in \sys.
When invoked by a subject, it doesn't necessarily mean that her PD will be completely and definitively erased.
In fact, it could be necessary to keep this data, for legal investigations for example.
In order to implement this, \sys assumes a model in which each data operator owns a public encryption key given to them by the authorities who keep the private key.
When PD is to be deleted, \sys will simply encrypt it using the public key, in this way, the data operator will not be able to access the data anymore, but the authorities will be able to decrypt it using their private key.
\section{Conclusion}
\label{conclusion}
We presented \sys, a GDPR-aware operating system, allowing every applications to be GDPR-compliant.
To this end, \sys introduces the concept of \textit{active data}.
Moreover, unlike current operating systems which are process-centric, \sys focuses on data.
We are currently implementing \sys following the semi-microkernel model, with Linux as a basis.
\section*{Acknowledgments}
This project is funded by several french research programs: ``Pack Ambition International'' - Région Auvergne-Rhône-Alpes 2021, MSH Lyon St-Etienne ``Projets thématique'' 2021 and Amorçage Europe 2022.


\end{document}